\documentclass[12pt,a4paper]{article}
\usepackage[cmex10]{amsmath}
\usepackage{float}
\usepackage{graphicx}
\usepackage{array}
\usepackage{color}
\usepackage{fixltx2e}

\begin{document}
\title{Ultracompact integrated polarizers using bent asymmetric coupled waveguides}

\author{P. Chamorro-Posada\\ 
Departmento de Teor\'{\i}a de la Se\~nal y Comunicaciones e\\
 Ingenier\'{\i}a Telem\'atica, Universidad de Valladolid,\\
 ETSI Telecomunicaci\'on, Paseo Bel\'en 15,\\
 E-47011 Valladolid, Spain.}

\maketitle

\begin{abstract}

A new type of integrated polarizer based on bent coupled optical waveguides is proposed.  The simplicity of the device geometry permits its realization using only basic standard fabrication steps.  A deep etched waveguide silicon nitride TE-pass implementation has been studied using a full vector 3D mode solver and beam propagation techniques.   The characteristics of the underlying physical mechanism bring in flexibility to the design of broadband devices robust against fabrication tolerances. The possibility of incorporating almost design-transparent polarizers in photonic integrated circuits is also discussed.

\end{abstract}

Polarization handling is a key issue in photonic integrated circuits (PICs) \cite{review}. Related devices include polarization beam splitters, polarization rotators and polarizers. Integrated polarizers are based on different propagation mechanisms that permit to discriminate either the TE or TM field by means of a strong attenuation on the crossed polarization.  The large polarization-dependent losses of the plasmonic modes can be exploited by the use of metals \cite{zhang14,li08,bai2017,sun16,sun17,ng12,xu15}, transition-metal oxides with metal-insulator transitions \cite{sanchez15}, graphene \cite{hao2015,yin2015} or transparent conducting oxides \cite{xu16}.  Purely dielectric polarizers include designs based on resonant tunneling structures \cite{azzam15},  shallowly-etched silicon-on-insulator ridge optical waveguides \cite{dai2010}, symmetric etching of silicon nanowires \cite{azzam14,wang10} or subwavelength gratings \cite{guan14,kim16}.

In this work, a new all-dielectric polarizer architecture is proposed. It is based on bent coupled waveguides \cite{chamorro17,chamorro18}.  These structures, originally proposed for the reduction of the radiation loss in improved-Q resonators, can perform strong differential losses for TE/TM polarizations.  We will consider the geometries of previous works \cite{chamorro17,chamorro18} based on the assumption of deep etched Si\textsubscript{3}N\textsubscript{4} rectangular waveguide with a SiO\textsubscript{2} cladding and a center operation wavelength of $\lambda=1.55$ $\mu$m. The width of transmission waveguides is $w=1$ $\mu$m and all the waveguides have a height of $h=300$ nm. The refractive index of silicon nitride is $1.9792$ and that of the oxide $1.4501$.   This can be regarded as a standard geometry for silicon nitride integrated circuits.  

\begin{figure}[htbp]
\centering
\begin{tabular}{cc}
\includegraphics[width=.45\linewidth]{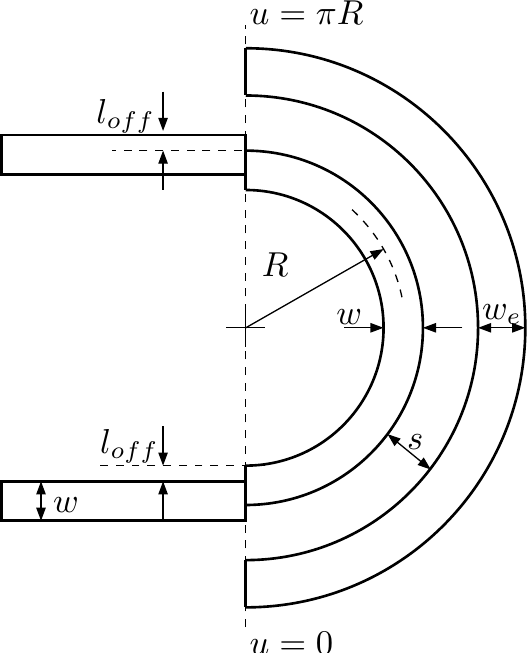}&
\includegraphics[width=.45\linewidth]{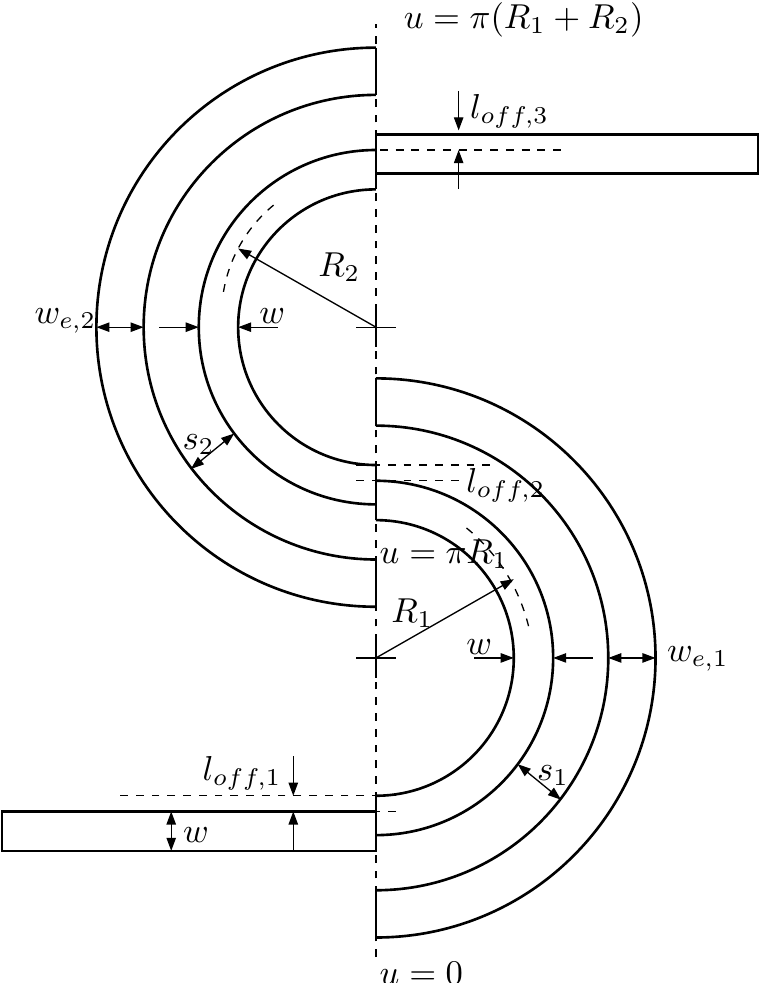}
\end{tabular}
\caption{Basic device geometry (left) and a example of a modified configuration (right).}
\label{fig::disenio}
\end{figure}

The waveguide under analysis will always support at least one TE-like and one TM-like propagating mode with very different group indices, modal profiles and bend losses.  Therefore, any optical device implemented in the PIC will exhibit a strong polarization-dependent response.  The placement of an input polarizer is a simple solution to this issue.  With the specifications detailed above, the bend losses are much larger for TM-polarized fields and devices will typically operate with TE optical signals.  The waveguide geometry used leads naturally to a TE-pass polarizer, which is precisely the one of interest for this implementation.  Alternative waveguide geometries will provide polarizers passing the verctically-polarized light with the same operation fundamentals.

\begin{figure}[htbp]
\centering
\begin{tabular}{c}
\includegraphics[width=.6\linewidth]{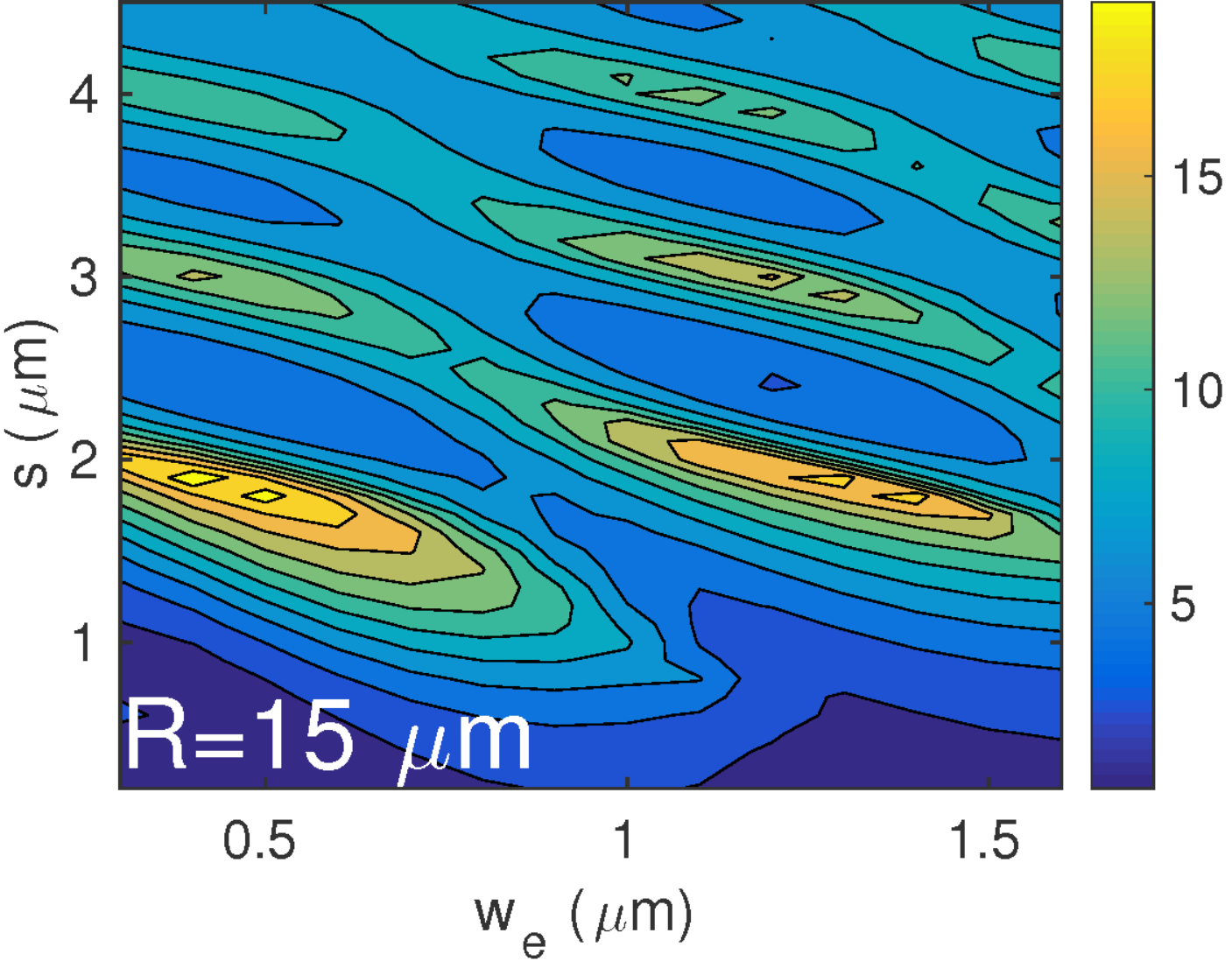}\\
\includegraphics[width=.6\linewidth]{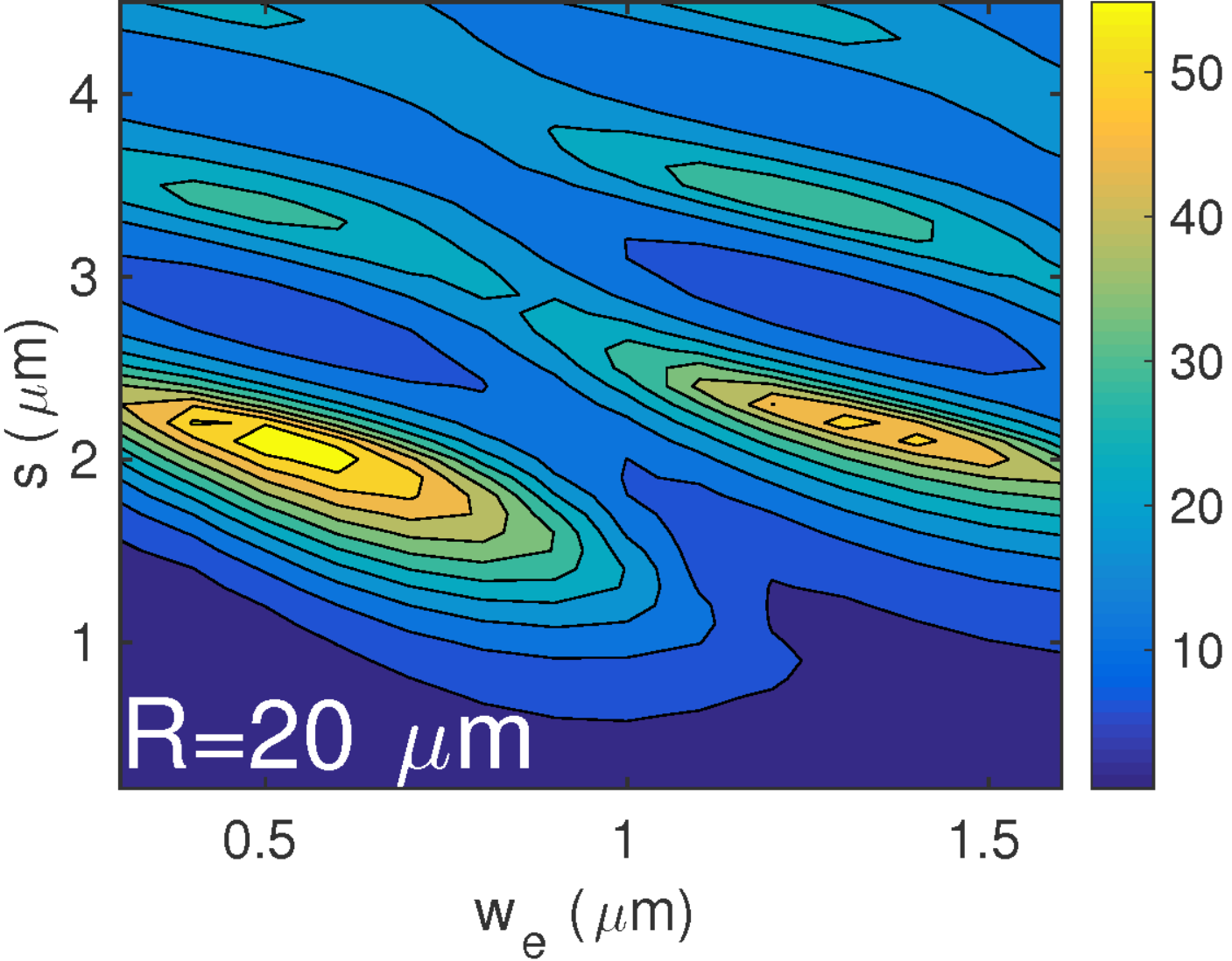}\\
\includegraphics[width=.6\linewidth]{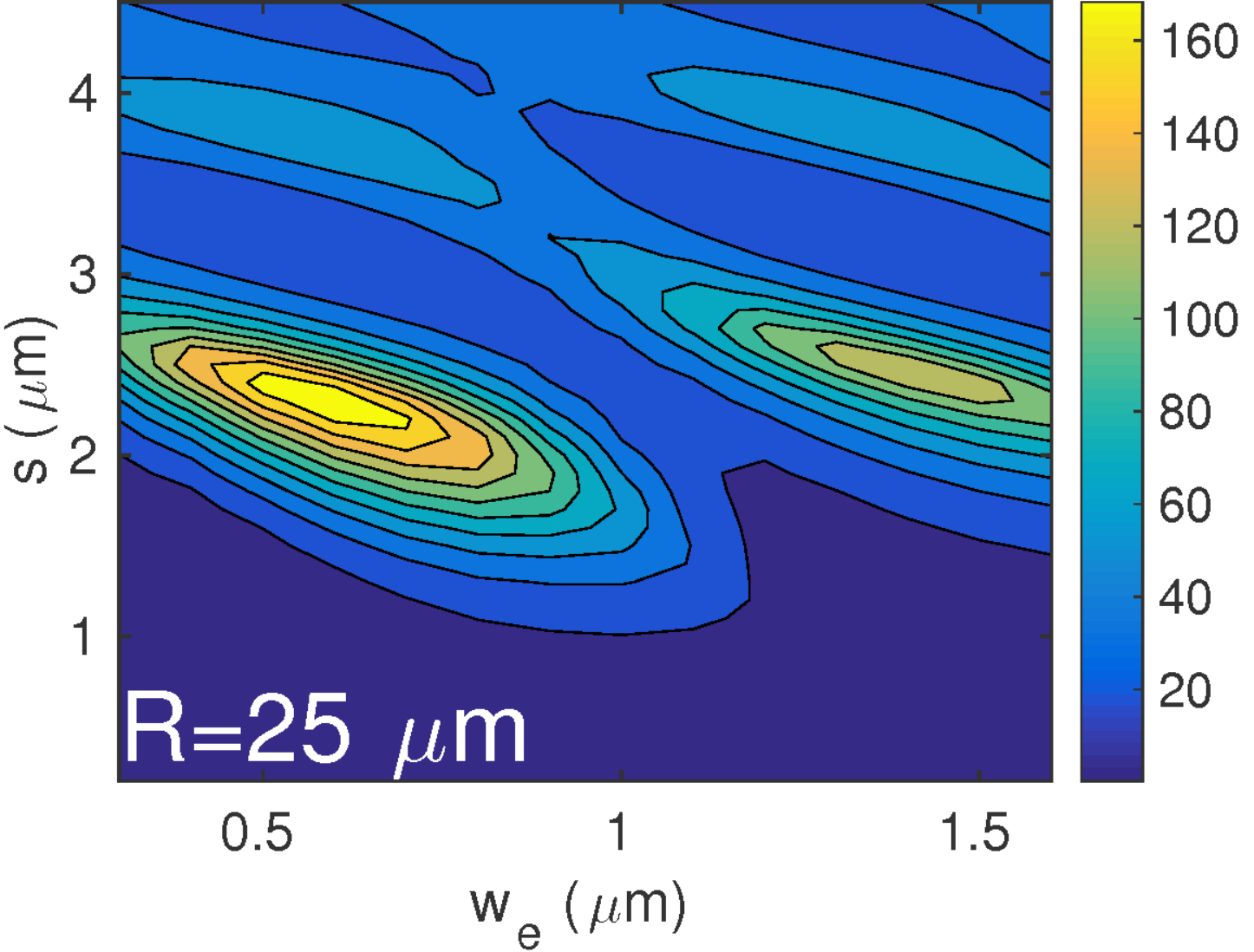}
\end{tabular}
\caption{Ratio of the imaginary modal index of the TE mode to that of the TM mode for different parameter sets $(s,w_e)$ and three values of the radius of curvature $R$.}
\label{fig::ni}
\end{figure}

\begin{figure}[htbp]
\centering
\includegraphics[width=\linewidth]{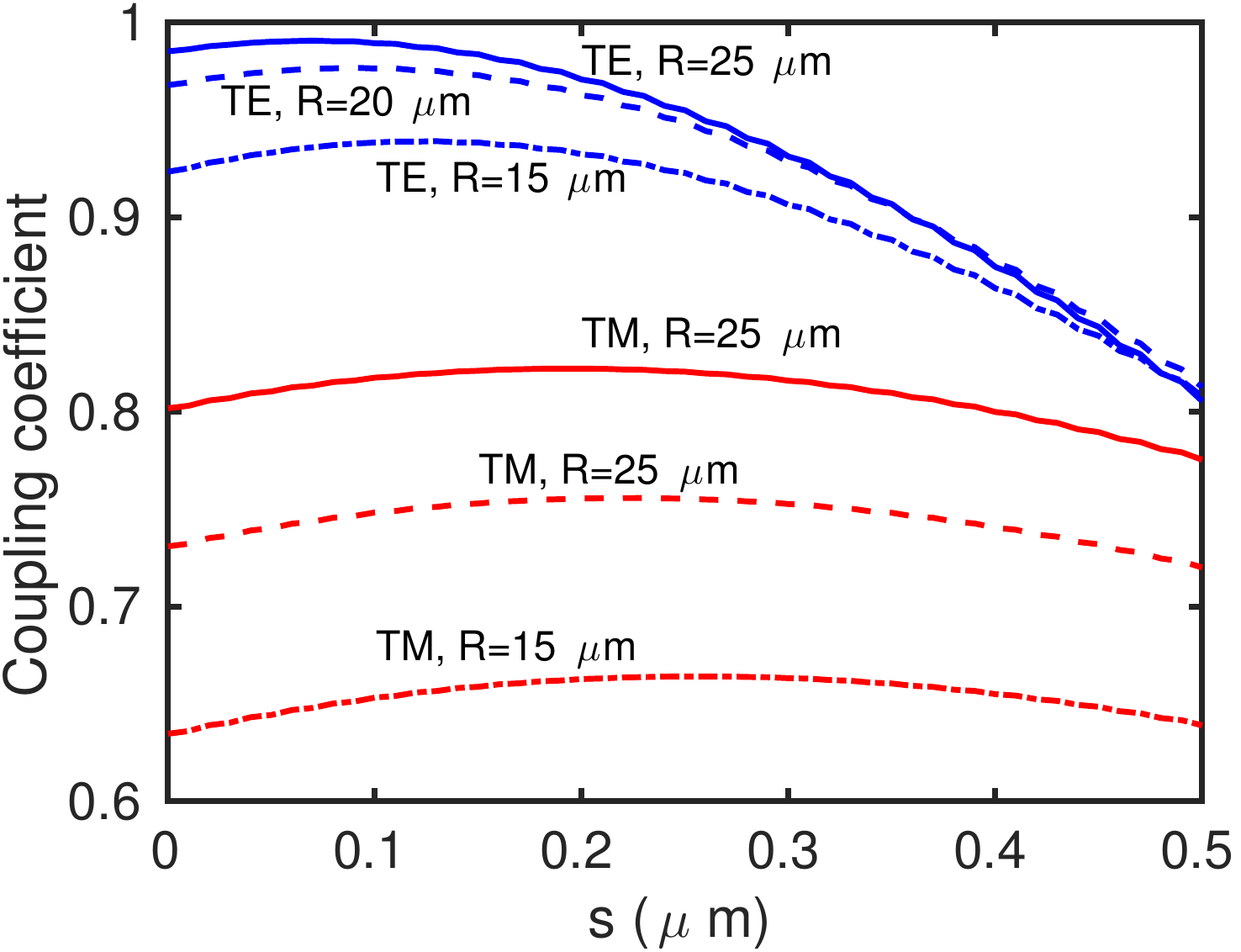}
\caption{Coupling loss between the straight and bent waveguide sections.}
\label{fig::loff}
\end{figure}

The device geometry is shown in Fig. \ref{fig::disenio} (left).  It consists of two coupled bent waveguides.  Optical signal transmission takes place in the interior waveguide, whereas the exterior arc is designed to control the radiation losses by adjusting the separation $s$ and the width of the second waveguide $w_e$.  Since all the waveguide structures involved in the design have a simple rectangular cross-section with the same height, the proposed devices can be fabricated using only basic standard steps.   Additional coupling losses arise due to the discontinuities at the input/output connections with the straight waveguides.  These are minimized by adjusting the lateral offset $l_{off}$ \cite{kitoh}.  The structure is very flexible and can be modified by using smaller-than-$\pi$ sections and/or connecting several of these sections as shown in Fig. \ref{fig::disenio} (right).

\begin{table}[htbp]
\centering
\caption{\bf Design examples}
\begin{tabular}{cccccc}
\hline
$R$ ($\mu$m) & $l_{off}$  (nm) & $s$ ($\mu$m) & $w_e$ (nm) & IL (dB) & ER (dB)\\
\hline
$15$         & $130$          & $1.9$         &  $400$       & $2.53$    & $34.86$   \\
$20$         & $90$          & $2.1$         &  $500$       & $0.75$    & $24.95$   \\
$25$         & $70$          & $2.3$         &  $600$       & $0.23$    & $18.46$   \\
\hline
\end{tabular}
  \label{tab::disenios}
\end{table}

An in-depth analysis of modal properties of curved asymmetric coupled waveguides used in our polarizers has been presented in \cite{chamorro18}.  The study is based on intensive 3D full-vector numerical computations and has permitted to identify the modes that remain predominantly confined in the transmission (inner) waveguide.  Figure \ref{fig::ni} displays the imaginary indices $n_i$ of these modes, relevant for the calculation of the radiation loss as calculated using the \emph{wgms3d} software package  \cite{krause} .  The propagation constant of the mode is $2\pi/\lambda_0(n_r+jn_i)$.  In order to visually determine the optimal operation conditions, the ratio of $n_i$ for the TE mode over that of $n_i$ for the TM mode over the design space $(s,w_e)$ is plotted. For this waveguide geometry, this ratio is always larger that $1$ and only TE-pass configurations are available.  For an inverted aspect ratio of the transverse section of the waveguide, a vertically polarized output could be obtained.  The optimal design regions (systematically close to the left lower corner of the plots) define broad regions that provide implementations that are robust against fabrication tolerances, since small deviations in the fabrication process will have little effect on the device performance.  The presence of multiple local maxima provides added flexibility to the design since alternative close-to-optical device configuration can be used in the case of existence of geometry restraints in the fabrication process. On the other hand, former studies \cite{chamorro17} show that these dispersion plots have very little wavelength dependence, providing the basis for broadband designs.

Silicon waveguides, with a higher refractive index contrast, permit the realization of comparable radiation losses with smaller curvature radii and, therefore, are expected to permit the implementation or more compact devices with similar performances. 

Fig. \ref{fig::loff} displays the coupling loss of TE and TM modes as a function of the lateral offset $l_{off}$ for three values of the radius of curvature $R$.  As expected, the coupling loss grows with the waveguide curvature.  Also, it is significantly larger for TM polarized fields.  This is due to the fact that a larger fraction of the total intensity remais outside the waveguide core for this polarization.  Our polarizers will benefit from this behavior of the coupling loss, that adds to the differential polarization-dependent attenuation in the curved coupled waveguides.

Table \ref{tab::disenios} summarizes the results of Fig. \ref{fig::ni} and Fig. \ref{fig::loff} for three polarizers consisting of a single $\pi$ rad section as shown in Fig. \ref{fig::disenio} (left).  For $R=15$ $\mu$m we obtain a very compact device with very large extinction ratio (ER=$34.86$ dB) at the cost of relatively high insertion loss IL=$2.53$ dB.  The insertion loss is largely reduced for R=$20$ $\mu$m, with IL=$0.75$ dB and still a very good value of ER.  For $R=20$ $\mu$m, the insertion loss is negligible, while the extinction ratio is still high.  This radius could already be close to that of the curved sections that are required in the inter-device connections of any PIC.  The modification of several of these curved sections, reducing its curvature radius and using the proposed scheme and, therefore, making them polarization selective can be used to exploit the properties of our proposal with minimal changes of an original circuit design.  This could provide a route to almost design-transparent implementation of the proposed integrated polarizers.   

\begin{figure}[htbp]
\centering
\begin{tabular}{c}
\includegraphics[width=.8\linewidth]{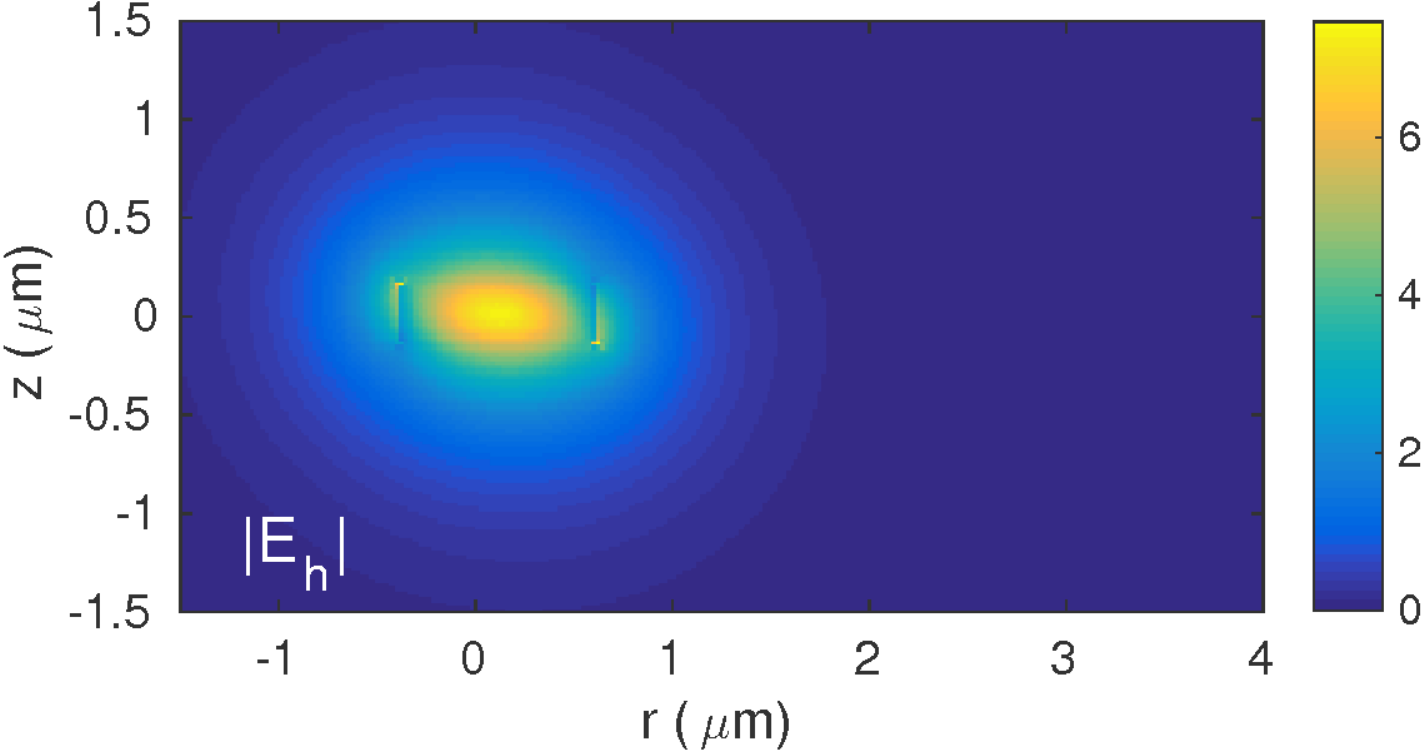}\\
\includegraphics[width=.8\linewidth]{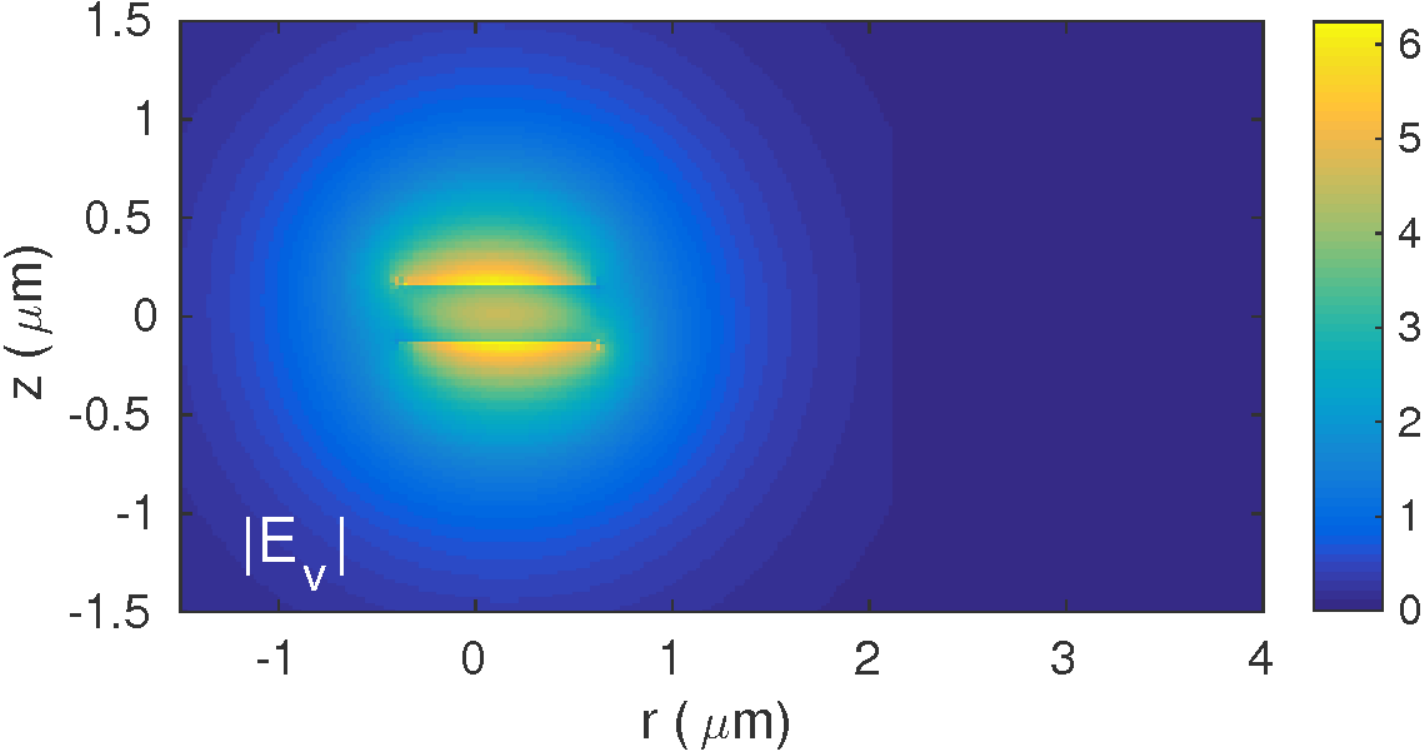}
\end{tabular}
\caption{Input transverse field in the BPM calculations.  $E_h$ and $E_v$ correspond, respectively, to the horizontally and vertically polarized components of the transverse electric field intensity.}
\label{fig::BPMentrada}
\end{figure}

\begin{figure}[htbp]
\centering
\begin{tabular}{c}
\includegraphics[width=.8\linewidth]{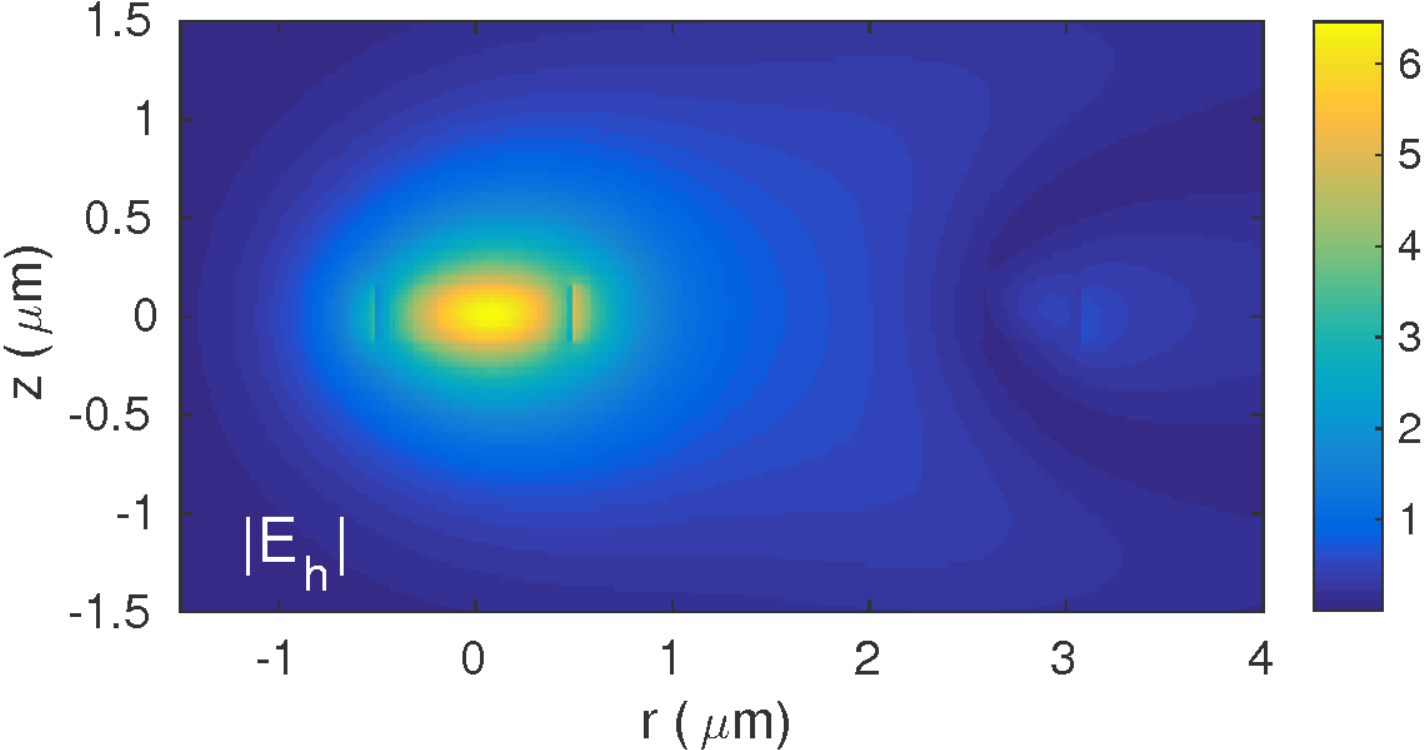}\\
\includegraphics[width=.8\linewidth]{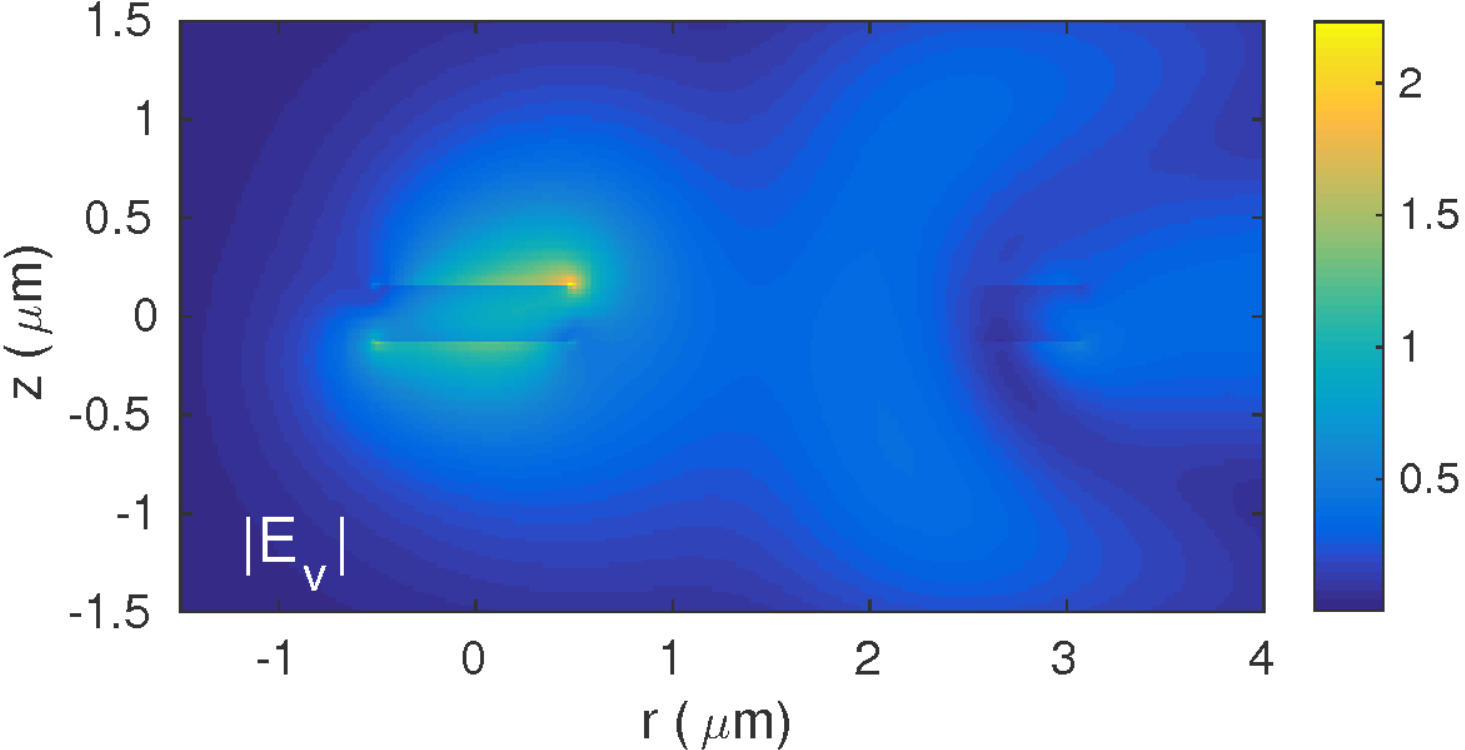}
\end{tabular}
\caption{Output transverse field in the BPM calculations.  $E_h$ and $E_v$ correspond, respectively, to the horizontally and vertically polarized components of the transverse electric field intensity.}
\label{fig::BPMsalida}
\end{figure}

\begin{figure}[htbp]
\centering
\begin{tabular}{c}
\includegraphics[width=.8\linewidth]{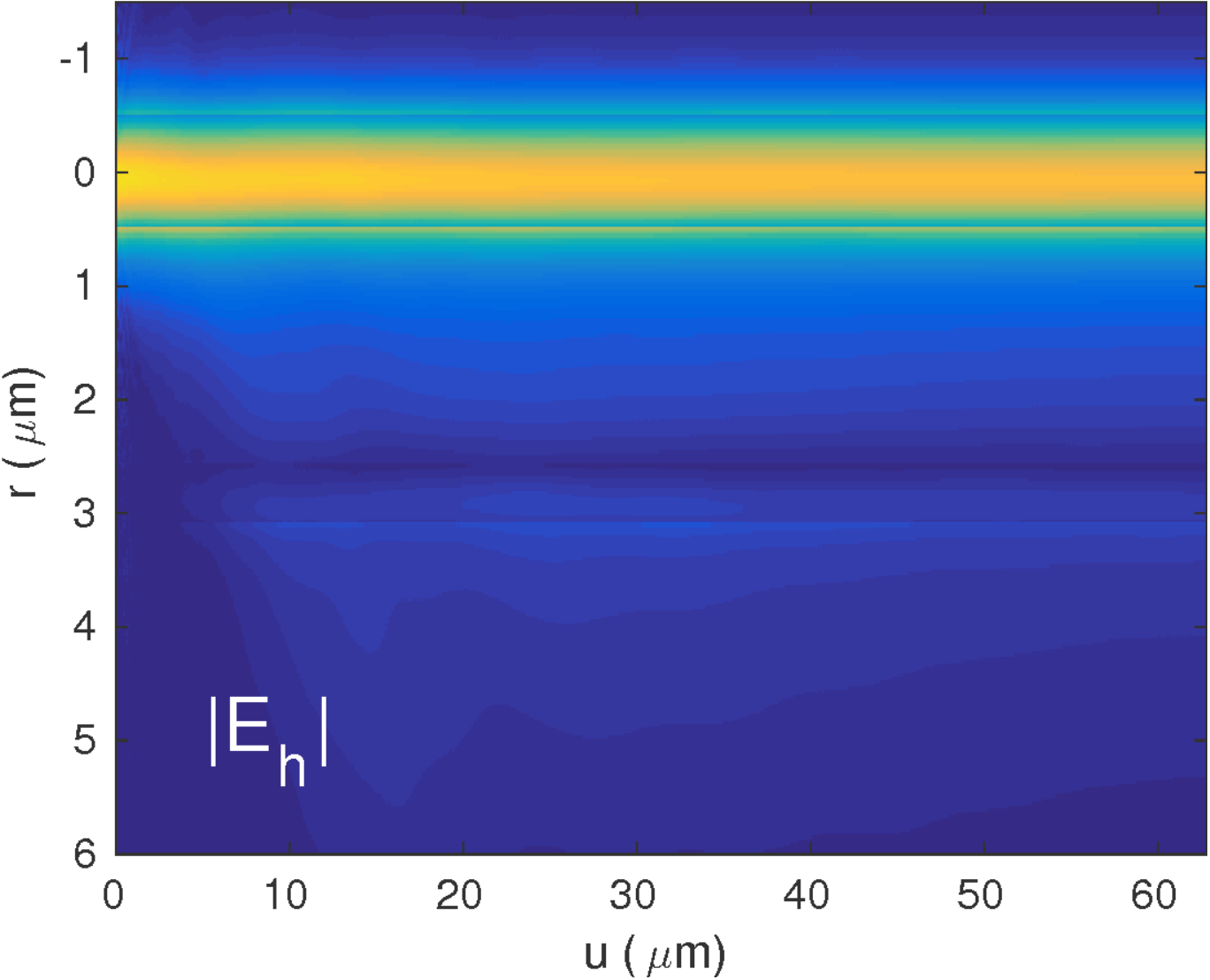}\\
\includegraphics[width=.8\linewidth]{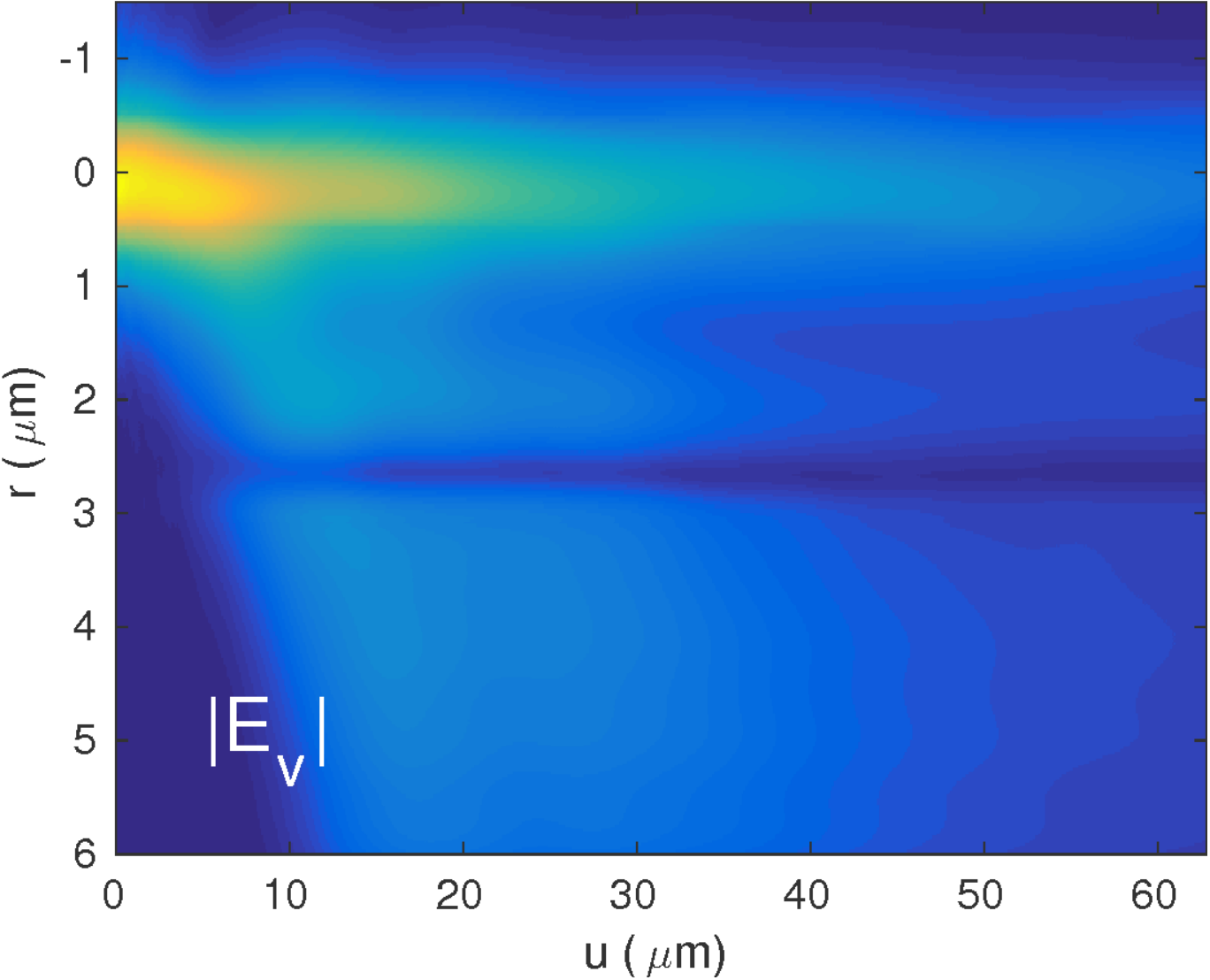}
\end{tabular}
\caption{Evolution of the transverse field amplitude in the horizontal plane crossing the center of the waveguides.}
\label{fig::BPMevolucion}
\end{figure}

The transmission properties of the designed polarizers have also been studied using a 3D full-vector finite-differences beam propagation method (BPM) \cite{lui}. Fig. \ref{fig::BPMentrada} shows the magnitude of the transverse input optical field consisting of equal-power contributions from the TE and TM polarized modes of the input straight waveguide.  The polarizer output can be observed in Fig. \ref{fig::BPMsalida}, after propagation along a $\pi R$-long curved coupled waveguide system with $R=20$ $\mu$m.  Fig. \ref{fig::BPMevolucion} displays the evolution of the amplitudes of the transverse horizontally and vertically polarized field components in the horizontal plane crossing at the waveguide center.  $r$ is the radial coordinate measured from the waveguide center and $u=R\phi$, where $\phi$ is the azimuthal coordinate. $z$ is the coordinate perpendicular to the plane defined by $r$ and $\phi$. The parameters used are $l_{off}=90$ nm, $s=2.1$ $\mu$m and $w_e=500$ nm.  The results are consistent with the modal analysis and they show a nearly TE-polarized output. Whereas the horizontally polarized field has survived almost intact to the propagation in the curved coupled waveguides, the vertically polarized component at the output is negligible, except for a spike close to the upper-right corner of the transmission waveguide, which has a vanishingly small contribution to the total output power.

In summary, a new type of all-dielectric integrated polarizer has been put forward.  The operation of the device is based on highly polarization dependent losses in coupled asymmetric curved waveguides.  The simple geometry of the design involve only the basic standard fabrication steps at the production stage.  Large extinction ratios and small insertion losses can be obtained in small-extension devices.  Alternatively, the proposed polarizers can be introduced at curved waveguides at different parts of a photonic integrated circuit with a minimal modification of the original design.  A detailed analysis of the performance of a device implemented in the silicon nitride platform has been carried out using highly accurate full-vector mode solving techniques and 3D vector beam propagation techniques.  For deep etched silicon nitride $1$ $\mu$m wide and $300$ tall waveguides, a device of size $46.4$ $\mu$m$\times 23.2$ $\mu$m provides $25.7$ dB of extinction ratio with an insertion loss of $0.75$ dB, while similar performances are expected with much more compact silicon devices.

This work has been was supported by Junta de Castilla y Le\'on,
Project No. VA089U16, and the Spanish Ministerio de Economía y
Competitividad, Project No. TEC2015-69665-R

\bibliography{polarizador.bib}{}
\bibliographystyle{unsrt}

\end{document}